\documentclass{nature}
\usepackage{color}

\title{The coevolution of overconfidence and bluffing in the resource competition game}

\author{Kun Li$^{1*}$, Attila Szolnoki$^{2}$, Rui Cong$^{3}$, Long Wang$^{1*}$}

\begin{document}

\maketitle

{
\small{
\begin{affiliations}
 \item Center for Systems and
Control, College of Engineering, Peking University, Beijing 100871, China
 \item Institute of Technical Physics and Materials Science, Centre for Energy Research, Hungarian Academy of Sciences, P.O. Box 49, H-1525 Budapest, Hungary
 \item Department of Automation and TNList, Tsinghua University, Beijing 100084, China
\item[$^*$]~Correspondence and requests for materials should be addressed to L.W. (email:
 longwang@pku.edu.cn); K.L.~(email: lk5622355@pku.edu.cn).
\end{affiliations}
}
}

\begin{abstract}
Resources are often limited, therefore it is essential how convincingly competitors present their claims for them. Beside a player's natural capacity, here overconfidence and bluffing may also play a decisive role and influence how to share a restricted reward. While bluff provides clear, but risky advantage, overconfidence, as a form of self-deception, could be harmful to its user. Still, it is a long-standing puzzle why these potentially damaging  biases are maintained and  evolving to a high level in the human society. Within the framework of evolutionary game theory, we present a simple version of resource competition game in which the coevolution of overconfidence and bluffing is fundamental, which is capable to explain their prevalence in structured populations.
Interestingly, bluffing seems apt to evolve to higher level than corresponding overconfidence and in general the former is less resistant to punishment than the latter. Moreover, topological feature of the social network plays an intricate role in the spreading of overconfidence and bluffing. While the heterogeneity of interactions facilitates bluffing, it also increases efficiency of adequate punishment against overconfident behavior. Furthermore, increasing the degree of homogeneous networks can trigger similar effect. We also observed that having high real capability may accommodate both bluffing ability and overconfidence simultaneously.
\end{abstract}

The emergence of overconfidence is a well-established bias in which a person's subjective confidence in self-assessment is greater than the objective accuracy of those judgments, especially when confidence is relatively high~\cite{Pallier2002JOGP}. In human societies overconfidence has been recognized in many different ways, such as overestimation of one's actual performance, over-ranking of personal achievement relative to others, and the excessive certainty regarding the accuracy of individual beliefs~\cite{Moore2008PR}. Although it is often blamed for hubris, market bubbles, financial collapses, policy failures and costly wars, overconfidence remains prevalent in our daily experience~\cite{Camerer1999AER,BABCOCK1992IR,Kampmark2006JOAS}. Such a bias can evolve due to the competition of alternative strategies and may contribute significantly to the increase of morale, ambition, resolve and persistence~\cite{Trivers2000EPOHRB,McKay2009BABS,West2011NATURE,Johnson2011NATURE}. Very high levels of core self-evaluations, a stable personality trait composed of locus of control, neuroticism, self-efficacy, and self-esteem, may also be related to the overconfidence effect~\cite{KOEHLER1995CP,Judge1997ROB}.

As a concomitant bias, bluffing, also named boasting or exaggeration, is a representation of something in an excessive manner~\cite{Demaree2004CE,Mittenberg2002JOCAEN}. The boaster is regarded as one who pretends to have distinguished qualities, but has not at all or to a lesser degree~\cite{Trivers2000EPOHRB,Li2014SR,Demaree2004CE}. Usually bluffing is not reliably distinguished from true ability~\cite{SMITH1976AB} and exists in different forms, such as amplifying achievements, deceiving others expectations by
magnifying emotional expressions~\cite{Demaree2004CE}. It is important to stress that the deception profile, including the appropriate levels of overconfidence and bluffing intensities, plays a decisive role in determining what an individual gets in resource competitions. Our specific interest here is to explore how such profiles develop due to an evolutionary process.

The application of realistic evolutionary rule, however, requires sanctioning of uncovered bluffing, which represents a sort of social norm of the population. In fact, the ability to develop and enforce social norms is probably one of the distinguishing features of the human species~\cite{Fehr04EHB}.
Several experiments and theoretical investigations have revealed that sanctions are able to create a sufficiently strong selective pressure to prevent cheating, which is necessary to stabilize human cooperation~\cite{Fehr02Nature,Rockenbach06Nature,PercSR12,Rand11NatureCom,Sigmund10Nature,Szolnoki2013PRX,Chen2014SER}.
Similarly, the deception behavior, regardless of self-deception (overconfidence) or other-deception (bluffing), might be controlled by centralized sanctions. Although third-party punishment can rectify peer biases caused by deception in cognition, system bias, which represents an inclination of the whole group, is beyond its reach. This system bias can be brought by social comparison bias~\cite{Heckman1998ECO}, for instance, where most of the members in a group believe that they are better (worse) than the average level of this group in certain aspects, which is apparently against the basic mathematical principles~\cite{Brown2000EJSP,Bandura1991JPSP}. Mandatory rules and many other factors can also bring about such system bias~\cite{Gans1986PH}.

Taking all the factors above into consideration, we explore how overconfidence and bluffing evolve within the framework of a spatial resource competition model. Here we follow the successful method of evolutionary game theory, which proved to be particularly efficient to explain the emergence and maintenance of cooperation~\cite{MaynardSmith82book,Hofbauer98book,Nowak04Science,Perc07EL,Hofbauer2003BOTAMS,Chen2012PRE,Fu2012SRE,Jin2014SRE,PercNJP06,Xia2014AMC,Chen2009PRE,Sui2015PLA,Li2015PRE}. We suppose that instead of strategies, players imitate personal profiles during the evolutionary process~\cite{Szolnoki11EPL}. More specifically, to adapt this concept to the present model, overconfidence and bluffing, considered as peer biases, could be the subject of imitation~\cite{Li2014SR}. The key element of the proposed model is both overconfidence and bluffing can evolve simultaneously, which influence a player's success to reach the desired resource. Furthermore, as another crucial point of our model, we suppose that uncovered bluffing will destroy the reputation of the related player. Accordingly, the mentioned player's overconfidence and bluffing intensities fall onto the minimal levels that are available in the actual population.

By using this simple concept, we find that the general bluffing level always evolves to a higher level than overconfidence. The application of sanctions, when the positive values of system bias reveal more possible conflicts between competitors, lower the overconfidence and bluffing levels remarkably. Beyond these observations we pay special attention onto the possible consequence of interaction topology.
It is well known that the spatial structure of interaction graph can influence significantly the evolutionary outcome of competing strategies in social dilemmas~\cite{Nowak92Nature,Nakamaru97JTB,Santos05PRL,Fu2008PRE,Szolnoki11EPL,Wu2011PRE,RCong12PLoSONE,Szolnoki2013PRE,ZWang13SR,Wang2012SRE,Chen2014FBN,Rand2014PNAS,Perc2013IF}. Motivated by this fact, we test different representative topologies and explore their consequence on the evolution of overconfidence and bluffing. We find that heterogeneity may boost bluffing and facilitate punishment against individuals' overconfidence, while increasing available neighbors of each player on homogeneous networks has triggered similar effect.

\section*{Results}
We start by presenting the stationary overconfidence level $f_{O}$ and bluffing level $f_B$ as a function of resource-to-cost ratio $r/c$, obtained on square lattice, as shown in Fig.~1(a). It suggests that increasing $r/c$ does not noticeably change $f_{O}$, but decreases $f_B$, especially when system bias $\delta$ is relatively large. Lifting $\delta$ also significantly reduces overconfidence level $f_{O}$ for moderate punishment probability $p$ ($p=0.5$). Note that  positive values of $\delta$ induce extra conflicts, and thus boost the chances of centralized sanctions. Therein it seems that the values of $r/c$ have little impact on the stabilization of overconfidence, regardless of whether punishment is rare or frequent. Meanwhile, boast behavior ($f_B$) slightly decreases as $r/c$ increases. Importantly, the results for regular random graph with $k=4$ are in accordance with those for translation invariant square lattice. Thus it seems that the structure of interactions does not play a prominent role as long as the average degrees $k$ are identical. The value of $k$, however, could play a decisive role on the evolution of deception profile. To explore this effect, we investigate the impact of $r/c$ on $f_{O}$ and $f_B$ under extreme conditions ($p=1,\delta=1$) on homogeneous networks with different $k$ values ($k = 4; 8; 16$). As shown in Fig.~1(b) overconfidence almost goes extinct irrespective of the values of $r/c$ and $k$, showing that enough sanctions can effectively reduce the general level of overconfidence. Meanwhile, $f_B$ drops sharply to a minimum value as $r/c$ ascends when $k=4$, in contrast to larger degrees as $k=8$ or $k=16$. In other words, having more available neighbors partially offsets the effect of punishment on boasters.

We next evaluate the impact of  probability of punishment $p$ and system bias $\delta$ on general overconfidence level $f_{O}$ and bluffing level $f_B$ (see Fig.~2). Besides homogeneous networks, summarized in Fig.~2(a) and (b), we also explored the possible impact of interaction heterogeneity by considering BA scale-free networks, shown in Fig.~2(c). To avoid additional effects we used the same average degree $\langle k \rangle=4$ used for random graph in Fig.~2(a).
It can be observed that at any given value of $\delta$, increasing punishment rate $p$ will slightly reduce both $f_{O}$ and $f_B$. Meanwhile, for any given $p$, both $f_{O}$ and $f_B$ drop with $\delta$ monotonously, signalling that $\delta$ plays a decisive role in restraining the deception behaviours (both overconfidence and bluffing). This behaviour is based on the fact that large $\delta$ ensures frequent conflicts between competing players, which will reveal their real abilities. In the other extreme case, negative $\delta<0$ parameter values inhibits conflicts, which results in a prompt fixation into a high overconfidence and high bluffing deception profile (this case is not shown in figures). Moreover, another common trait of color maps independently of the applied topologies is that $f_B$ always evolves to a higher level than the corresponding $f_{O}$, highlighting that natural selection provides higher bluffing level than overconfidence when other factors equal. Furthermore, the comparison of Fig.~2(a) and Fig.~2(c) illustrates that network heterogeneity can apparently elevate average bluffing intensity, $f_B$. It also illustrates that the heterogeneity of interaction topology helps to restrain overconfidence for relatively large $\delta$ values. Interestingly, increasing $k$ of homogeneous networks is capble to lift bluffing level $f_B$ while overconfidence $f_{O}$ is slightly reduced (see also Fig.~1(b)).

For better understanding the possible influence of sanctioning mechanism on the evolution of deception profile $(\alpha, \beta)$, we monitor the time evolution of $\alpha$ and $\beta$ values on a square lattice without and with punishment (shown in Fig.~3(a) and in Fig.~3(b), respectively). Fig.~3(a) shows how the probability distribution of $f(\alpha, \beta)$ pairs evolve in time in the absence of punishment, when only imitation of deception profiles is possible. It can be observed that the small $\beta$ values die out  first, signalling that boast is most favored by natural selection. Later, when only large $\beta$ values are present, those players become more successful who apply higher $\alpha$ values. As a result, the whole population will be trapped into a
large $(\alpha, \beta)$ pair after sufficiently long relaxation ($t=100000$ MC steps). In fact, once fixation occurs the evolutionary process stops. Here, $f_{O}$ and $f_B$ can then be determined by means of averaging over the final states that emerge from different initial conditions. As we conclude, a high $\alpha$$-$high $\beta$ combination survives when there exist only imitations, which is in accordance with our previous observations~\cite{Li2014SR}. However, fixation never happens when sanction determines the evolution (see Fig.~3(b)). In the early stage almost half of the population is punished, hence low $\alpha$$-$low $\beta$ combinations will form the majority of $f(\alpha, \beta)$ distribution. Later, as time passes, a dynamic balance emerges between low $\alpha$$-$low $\beta$ combinations and a moderate $\alpha$$-$high $\beta$ pairs. The specific position of the latter depend on the actual values of $\delta$ and $p$ parameters. In general, the punishment plays a ``shunting'' role here, undermining the stabilization of overconfidence and bluffing in the whole population. Importantly, these results hold for any homogenous networks besides square lattice. For strongly heterogeneous networks, sometimes more than one $\alpha-\beta$ pair can survive around strong hubs even without punishment, which is in agreement with related works where other player-specific profiles evolved~\cite{Szolnoki11EPL,Szolnoki2013PRE}.

After realizing the significant impact of sanctions on the evolution of deception profile, next we are interested in the  targets of such punishments. More precisely, we wonder whether the real inferiors' deception profiles are minimized on homogeneous networks with different $k$ values. For this reason we measure separately the average real capability of those players who are punished and those players who are not. The ratio of their averages is denoted by $R_{ability}$. Similarly, we also measure the average payoff of the mentioned subclasses, and their ratio is denoted by $R_{payoff}$. These ratios are depicted in Fig.~4(a) for different random regular networks, where we gradually increase the degree $k$. Apparently, $R_{ability}<1$ indicates that on average, players having lower real $\gamma$ abilities are punished more frequently. At the same time, $R_{payoff}<1$ values highlight that the mentioned small-$\gamma$ group benefit less than their higher ability opponents. Increasing the degree of nodes, both $R_{ability}$ and $R_{payoff}$ raise unambiguously, showing that enhancement of connections narrows the real capability- and payoff-gap between the punished players and those who are not punished. In other words, punishment is directed principally towards who are really weak, but this selective impact is gradually weakened as each one has more neighbors. Furthermore, for a deeper insight, it is worth studying the influence of real capability on the evolution of overconfidence and bluffing. Note that the real ability $\gamma$ of each player remains unchanged during updating. We mark by $R_O([a,b])$ the ratio of the average overconfidence level of those individuals whose $\gamma$ values are in the $[a,b]$ interval compared to the whole population. For simplicity, we divided the $[0,\gamma_{max}]$ interval into $10$ subclasses. Similarly, $R_B([a,b])$ denotes the ratio of bluffing level for the same subpopulation. Our results for $k=4$ random regular graph are summarized in Fig.~4(b). The plot suggests clearly that both $R_O$ and $R_B$ ascend with $\gamma$, and exceed the ratio 1 once $\gamma>0.5$. Note that homogeneous networks with other $k$ values show similar tendency. Thus we conclude that players with high ability are inclined to evolve to a higher state of both overconfidence and bluffing because they have a higher chance to collect resource without conflict. Furthermore, if conflict is inevitable and competitors should reveal their real abilities then the mentioned players still have a higher chance to win.

Lastly, it is instructive to investigate the impact of upper limits $\alpha_{max}$ and $\gamma_{max}$ on the evolution of $f_{O}$ and $f_{B}$ values.
By keeping $\gamma_{max}=\beta_{max}=1$, $\alpha_{max}>1$ means that excessive overconfidence intensity is allowed for competitors.
$\gamma_{max}>1$, when $\alpha_{max}=\beta_{max}=1$, however, implies that real abilities of players are significantly higher compared to the changing $\alpha$ or $\beta$ values. We note that $\beta_{max}>1$ is not taken into consideration, for extravagant boasting could be easily recognized from real facts.
For appropriate comparison, $f_{O}$ is normalized, $f_{O}^{norm}=f_{O}/\alpha_{max}$, when $\alpha_{max}>1$ is applied. As demonstrated in Fig.~5(a), the possibility of sanctioning results in drastic reductions in the normalized overconfidence level $f_{O}^{norm}$ as $ \alpha_{max}$ is increased. It suggests that punishment can effectively restrain excessive overconfidence, but is unable to decrease bluffing level significantly. However, without punishment ($p=0$), raising $\alpha_{max}$ gives rise to intensive conflicts that help competitors to recognize others' real capabilities. Therefore, $f_{O}^{norm}$ and $f_{B}$ monotonously decrease with $\alpha_{max}$, and finally converge to 0.5, which equals to the initial value of average bluffing intensities.
We stress that the results presented in Fig.~5(a) are robust and remain valid if we use other interaction topologies.
Increasing $\gamma_{max}$ drives the evolution toward ``neutral drift'' because peer biases, such as overconfidence and bluffing, become second-order important in resource competitions when real abilities dominate. Importantly, however, $f_{O}$ and $f_{B}$ may fluctuate heavily in heterogeneous networks, showing that the existence of strong hubs might influence significantly the evolution both overconfidence and bluffing.

\section*{Discussion}
In summary, we have investigated how overconfidence and bluffing co-evolves within the framework of a resource competition game.
It is a well recognized fact that when confidence is relatively high then the whole population fall victim easily into overconfidence, which is considered to be the most ``pervasive and potentially catastrophic'' of all the cognitive biases by some psychologists~\cite{Pallier2002JOGP,KOEHLER1995CP}.
Counterintuitively, this ``erroneous'' psychology can maximize individual fitness in many situations, leading to its prosperity in human society. Meanwhile, the existence of bluffing behavior, sometimes unable to be detected, usually leads to ambiguity in one's perception about other's real ability. Our previous study highlighted that bluffing promotes overconfidence and they both stabilize at a high level when evolution is limited via imitation without the chance to reveal competitors' real abilities~\cite{Li2014SR}. However, the ability to develop and enforce social norms is probably one of the most characteristic feature of human species~\cite{Fehr04EHB}. Motivated by this fact  we propose an evolutionary which combines sanction mechanism with the clebrated rule of ``imitating the better''~\cite{Zimmermann05PRE}. Punishment here, instead of reducing individuals' real income, is only reduced to their deception behaviors, including both self-deception (overconfidence) and other-deception (bluffing). It is a key point of our model that these two mechanisms, which may determine a player's success, can coevolve. Furthermore, except the deception profile, the system bias describing the group inclination towards extra conflicts is also considered. Accordingly, system bias can be treated as integral effect, caused by all the other factors, to stimulate conflicts ($\delta>0$) or to inhibit conflicts ($\delta<0$) between competitors. In addition, punishment is not certain to occur, but happens with probability $p$ here. Lastly, we stress that we have tested different interaction topologies to explore the possible consequences of structured population.
All these details make our model more realistic.

Our extended model gives deeper insight to previous findings~\cite{Li2014SR}. As shown in Fig.~2, overconfidence and bluffing have essentially the same changing tendency
irrespective of $p$, $\delta$ and topological properties. It is in accordance with previous observation that bluffing promotes overconfidence. There is, however, a significant difference, when both side of deception can coevolve. Namely, boasting seems more stable than the fatal psychology of overconfidence because individuals can take advantage of bluffing immediately. As a consequence, eliminating boast behavior requires more intensive sanction mechanism to work. We also find that increasing heterogeneity or average degree of the interaction networks significantly promote bluffing, and simultaneously increase efficiency of adequate punishment (when $p$ and $\delta$ are large) against overconfident behavior. More importantly, this third-party punishment prominently limits  overconfidence of excessive intensity. Intriguingly, high capability of an elite might induce high level of his deception profile, which lies in the fact that elites hardly fail in the conflicts.

In conclusion, for better understanding the intricate relation between overconfidence and bluffing, we have proposed a more realistic model in which the individual deception profile coevolve. Overall, both social norms and  topological properties of interaction networks have substantial influence on the evolution of these ``peer biases''. We hope that these observations will motivate further research aimed at promoting our comprehension of the evolution of these ``erroneous'' but sometimes meaningful inclinations.

\section*{Methods}
The traditional setup of an evolutionary game assumes $N$ players occupying vertices of an interaction graph.
Our basic model is a resource competition game (RCG) in which neighbors compete for resources and their success is based on how convincingly they claim for it. Without loss of generality, an individual $i$ is characterized by a time-independent real capability $\gamma_{i}\in [0,\gamma_{max}]$, and an evolving overconfidence intensity $\alpha_i \in [0,\alpha_{max}]$, and bluffing intensity $\beta_i \in [0,\beta_{max}]$. Here $ \gamma_{max}$, $ \alpha_{max}$, $\beta_{max}$ values represent upper limits of corresponding properties of the whole population. Unless stated, $\gamma_{max}=\alpha_{max}=\beta_{max}=1$.
While the real capacity $\gamma_i$ is fixed and unalienable feature of each players, $\alpha_{i}$ represents the actual overconfidence state (OS), a perception error about self-ability. Similarly, $\beta_{i}$ characterizes the bluffing state (BS) of the player that helps to over-represent abilities towards competitors. In particular, $i$ believes he/she owns a ``self-perceived capability'' $ k_{i}$ as:
       \begin{equation}
                 k_{i}=\gamma_{i}+\alpha_{i} \,,
       \end{equation}
while his/her ``displaying capability'' $ m_{i}$ is observed as:
              \begin{equation}
                 m_{i}=\gamma_{i}+\beta_{i} \,.
               \end{equation}
Supposing a resource $r$ is potentially available to neighboring individuals that claim it. If neither of them claims then the resource remains unused. If only one individual makes a claim, then it acquires the resource and gains fitness $r$ while the other gains nothing.
When, players $i$ and $j$ both claim for this resource, a RCG takes place. In the latter case each individual pays a cost $c$ due to the conflict between them, and the one who has higher real capability acquires the resource. In this model, the recognition ability of each player is also influenced by a uniform system bias $\delta$, which allows us to control the intensity of conflicts between competitors. Summing up, a player $i$ facing with player $j$ gains a payoff $P_{ij}$ that can be calculated as follows:
     \begin{itemize}
                          \item (1)~~If $k_{i}>m_{j}-\delta$ and $k_{j}<m_{i}-\delta$, player $i$ claims but player $j$ does not, thus $P_{ij}=r$.
                          \item (2)~~If $k_{i}<m_{j}-\delta$, player $i$ will not claim and remains empty handed, $P_{ij}=0$.
                          \item (3)~~If $k_{i}>m_{j}-\delta$ and $k_{j}>m_{i}-\delta$, a conflict emerges between players $i$ and $j$ when they have to reveal their real capabilities which determine what they get:
                     If  $\gamma_{i}>\gamma_{j}, P_{ij}=r-c$; If $\gamma_{i}<\gamma_{j}, P_{ij}=-c$.
     \end{itemize}
Here parameter $\delta$ represents a uniform group inclination how to handle possible conflicts: for positive $\delta>0$ values group members are motivated to ``open their cards'' impulsively and bravely, and thus more conflicts take place. In case of $\delta<0$, however, conflicts are avoided because all players in the group are excessively cautious.

Initially each player $i$ is assigned by random $ \gamma_{i}$, $ \alpha_{i}$ and $\beta_{i}$ values. The situation that two values are equal is not taken into consideration.
In stark contrast to our preliminary work~\cite{Li2014SR} in the extended model both $\alpha_i$ and $\beta_i$ can coevolve, which influence dramatically a player's success in resource competition. During an elementary Monte Carlo (MC) step a randomly selected player $i$ collects its payoff $P_{i}$ by playing RCG with all $k_i$ neighbors, where $k_i$ represents the degree of player $i$ in the interaction graph. The total payoff of player $i$ is
  \begin {equation}
    P_{i}=\sum_{j \in \Omega _{(i)}} P_{ij}\,,
   \end{equation}
where $ \Omega _{(i)}$ represents all players in $i$'s neighborhood. Subsequently, a randomly chosen neighbor $j$ acquires its payoff $P_j$ in a similar way.

As we noted, a crucial point of the evolution that players may change their deception profile to collect more resources. In particular, if a player $i$ looses a conflict against player $j$ then his/her extreme overconfidence and bluff levels are revealed, hence player $i$ is punished with probability $p$. As a result, the ($\alpha_i$; $\beta_i$) values are reduced to the minimum levels of the whole population. Otherwise, player $i$ adopts the deception profile of a randomly selected neighbor $j$ with the probability $ W=W(P_{j}-P_{i})$. And thus
\begin{equation}
 \left\{
   \begin{array}{ll}
     k_{i}>m_{j}-\delta,k_{j}>m_{i}-\delta,\gamma_{i}<\gamma_{j}: \left\{
                   \begin{array}{ll}
                    p: \alpha_i=\varepsilon_\alpha,\beta_i=\varepsilon_\beta  \\
                   1-p: W(P_{j}-P_{i})={\Big(1+exp[ (P_{i}-P_{j})/K    ]\Big)}^{-1}
                   \end{array}
                 \right.
\\
     \textrm{Otherwise } W(P_{j}-P_{i})={\Big(1+exp[ (P_{i}-P_{j})/K    ]\Big)}^{-1}\,\,,
   \end{array}
 \right.
 \end{equation}
 where $\varepsilon_\alpha$ and $\varepsilon_\beta$ represent the minimum overconfidence and bluffing intensity respectively. Parameter $K$ characterizes the level of uncertainty in deception profile adoption~\cite{Szabo07review}. Without loss of generality we use $K=0.1$, but qualitatively similar results can be obtained for other $K$ values. Importantly, since the profile consists of two parameters, two independent random numbers are drawn to enable uncorrelated imitation of $\alpha_i$ and $\beta_i$ values, as it was suggested in Ref~\cite{Szolnoki2013PRE}.

 The presented simulation results were obtained using different interaction graphs, such as square lattice with periodic boundary conditions, regular random graph with different degrees, and the Barab\'{a}si-Albert (BA) scale-free graph~\cite{Barabsi99science}. The latter is served to explore the possible consequence of heterogeneities. In accordance with the random sequential update, each full MC step, which consists of $N$ times of repeated elementary steps, gives a chance on average once to update individual deception profiles.The typical system size contains $N=10^4-10^5$ nodes
 and the stationary frequencies are determined by averaging over $10^4$ MC generations in the stationary state after sufficiently long relaxation times.
 The stationary state is considered to be reached when the average of the overconfidence level $f_O$ (the stable average values of $\alpha$) and bluffing level $f_B$ (the stable average values of $\beta$) no longer change in time. We have averaged the final outcome over 50 independent initial conditions.

\begin{addendum}
 \item The authors are supported by NSFC (Grants 61375120 and 61533001) and by the Hungarian National Research Fund (Grant K-101490).
 \item[Author Contributions]K.L., A.S., R.C. and L.W. performed analyses, discussed the results, and contributed to the text of the manuscript.
 \item[Competing Interests] The authors declare that they have no competing financial interests.

\end{addendum}

\begin{figure}
\caption{Stable overconfidence level $f_{O}$ and bluffing level $f_{B}$  as a function of resource-to-cost ratio $r/c$ for different values of $p$ and $\delta$ on homogeneous networks as indicated in the legends. Data presented in Panel~(a) are obtained on square lattices with periodic boundary conditions, while results depicted in Panel~(b) are obtained on regular random graphs with different values of degree ($k=4,8,16$) when optimal punishment is applied ($p=1$, $\delta=1$). Other parameters: $\gamma_{max}=\alpha_{max}=\beta_{max}=1$.}
\end{figure}

\begin{figure}
\caption{Color maps depicting the overconfidence level $f_O$ (left column) and the bluffing level $f_{B}$ (right column) on the punishment probability ($p$) - system bias ($\delta$) plane.
Data presented in Panel~(a) are obtained on regular random graph with $k=4$. Panel~(b) are obtained on regular random graph with $k=8$, while results depicted in Panel~(c) are obtained on BA scale-free network with $\langle k \rangle =4$. Note that $\delta<0$ immediately leads to $f_O\rightarrow1$ and $f_B\rightarrow1$ regardless of the applied topology (not shown). Other parameters: $r/c=2.5$, $\gamma_{max}=\alpha_{max}=\beta_{max}=1$. }
\end{figure}

\begin{figure}
\caption{Time evolution of the $\alpha-\beta$ profile, as obtained on square lattices (a) without punishment ($p=0$) and (b) with punishment ($p=1$).
From top left to bottom right we have presented the temporal distribution of $(\alpha,\beta)$ pairs at different MC steps, as indicated. The comparison illustrates that punishment undermines the fixation of overconfidence and bluffing. Other parameters: $\gamma_{max}=\alpha_{max}=\beta_{max}=1$, (a) $r/c=3$, $\delta=0$; (a) $r/c=3$, $\delta=1$.}
\end{figure}

\begin{figure}
\caption{Some representative ratios plotted in histogram forms. Panel~(a) depicts $R_{ability}$ and $R_{payoff}$ ratios as a function of degree $k$ on regular random networks. Here $R_{ability}$ ($R_{payoff}$) represents the ratio of the average real capability (the average payoff) of those individuals who are punished to those who are not punished. Panel~(b) depicts $R_{O}$ and $R_{B}$ as a function of real ability interval on regular random networks with $k=4$. Here $R_{O}$ denotes the ratio of the average overconfidence level of individuals whose real capability fits in corresponding interval of $f_O$. In a similar fashion, $R_{B}$ represents the ratio of the average bluffing level of individuals whose real capability belong into the corresponding interval of $f_B$. Other parameters: $p=1$, $\delta=0.8$, $\gamma_{max}=\alpha_{max}=\beta_{max}=1$.}
\end{figure}

\begin{figure}
\caption{Stationary overconfidence level $f_{O}$ and bluffing level $f_{B}$ as a function of upper limit of overconfidence intensity $\alpha_{max}$ (panel~(a)) and in dependence of the upper limit of real capability $\gamma_{max}$ (panel~(b)). Panel~(a) depicts the normalized overconfidence level $f_{O}^{norm}$ and bluffing level $f_{B}$ as a function of $\alpha_{max}$ without punishment ($p=0$) and with punishment ($p=0.3$) on square lattices, where $f_{O}^{norm}=f_{O}/\alpha_{max}$. Panel~(b) depicts $f_{O}$ and $f_{B}$ as a function of $\gamma_{max}$ on square lattice and BA network when $p=0.9$. Error bars indicate the standard deviations, which are almost invisible in homogeneous network. Other parameters: $\delta=0.2$, $r/c=2$.}
\end{figure}

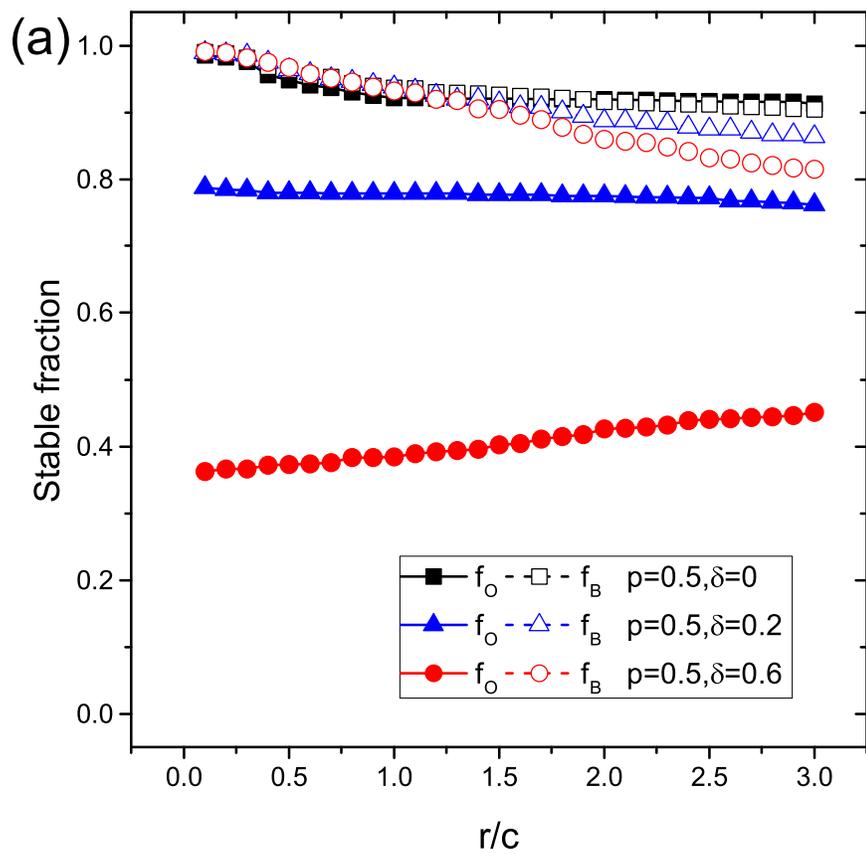
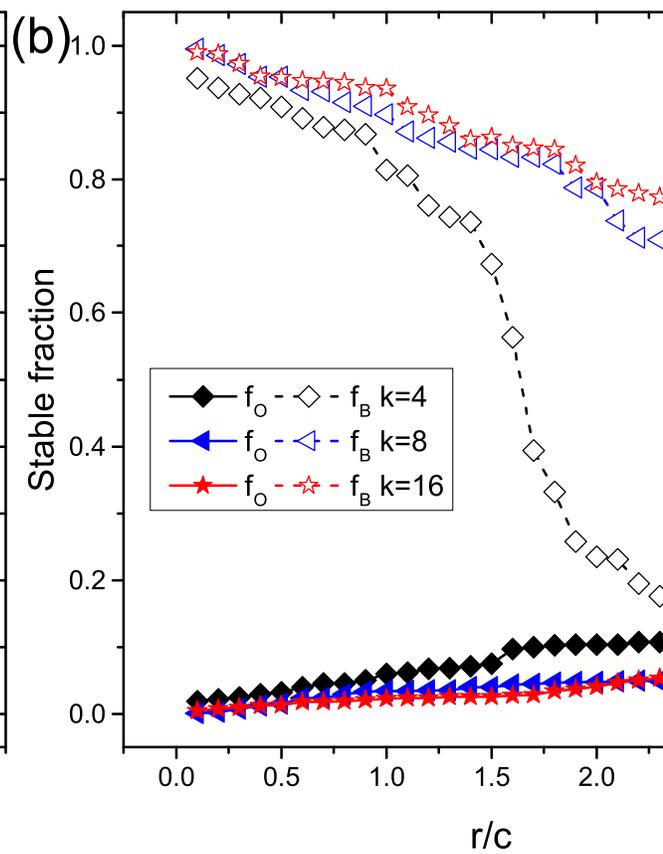
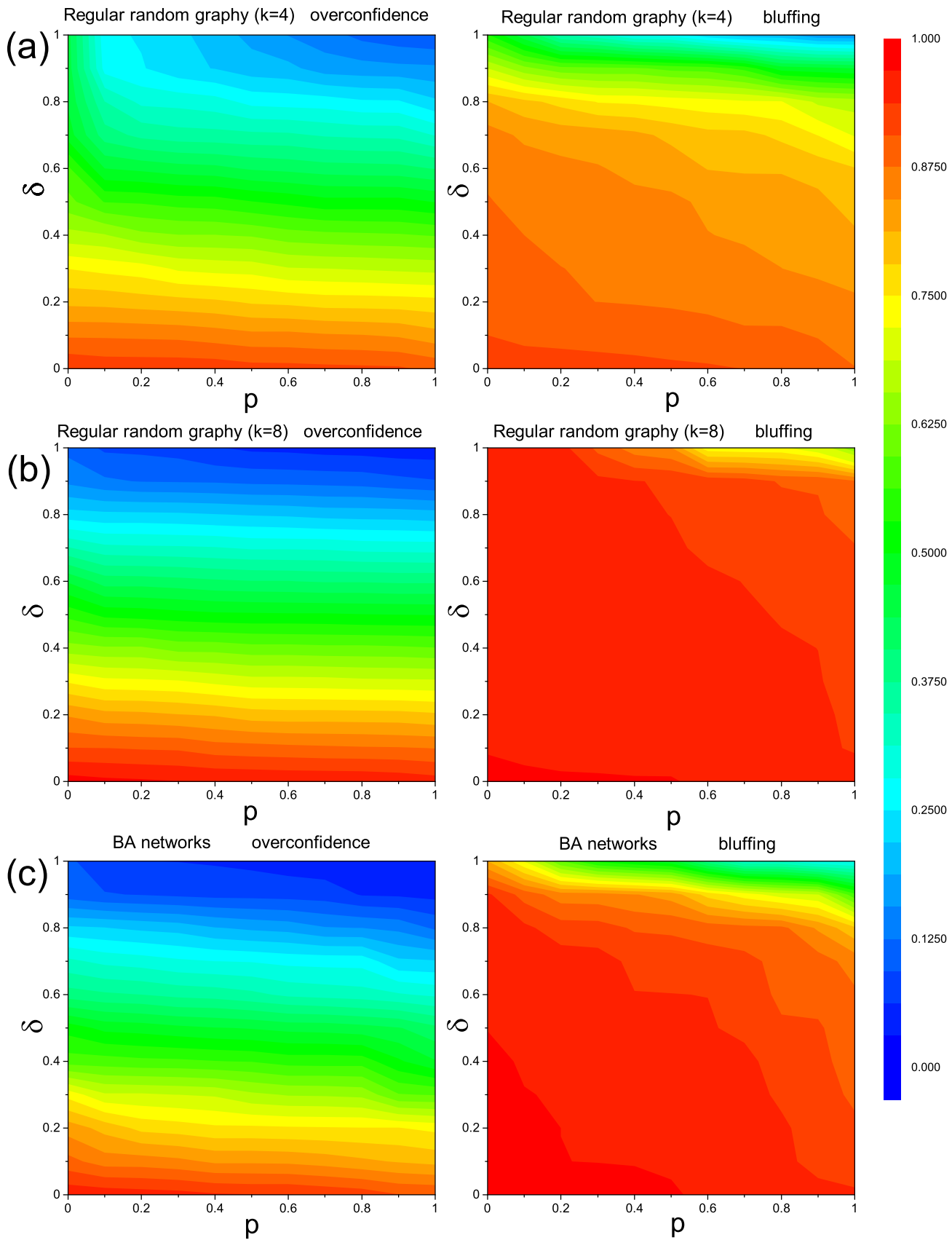
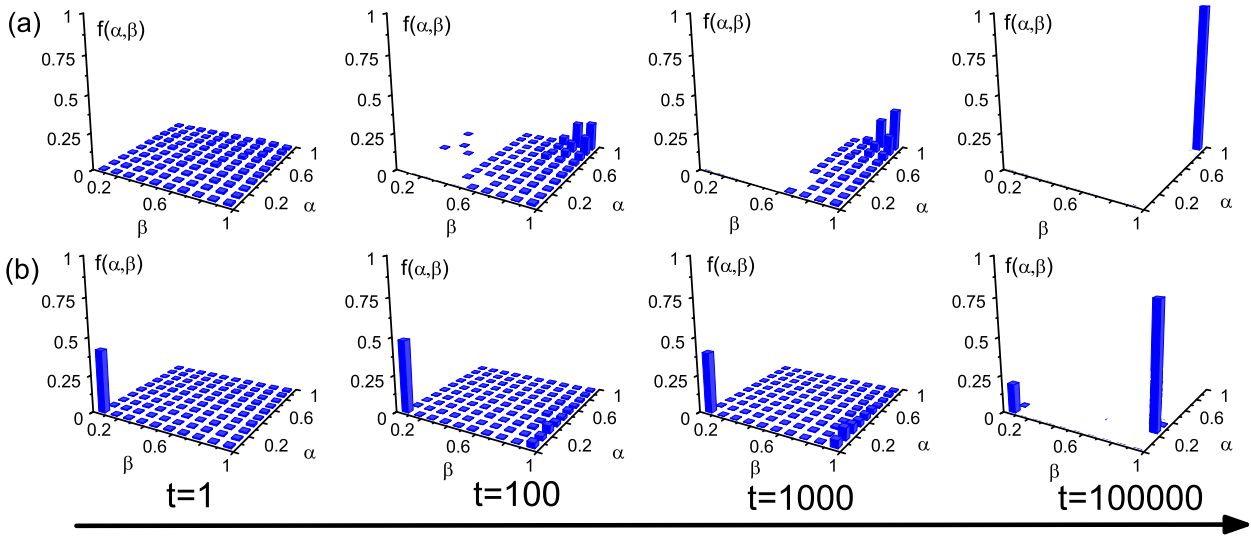
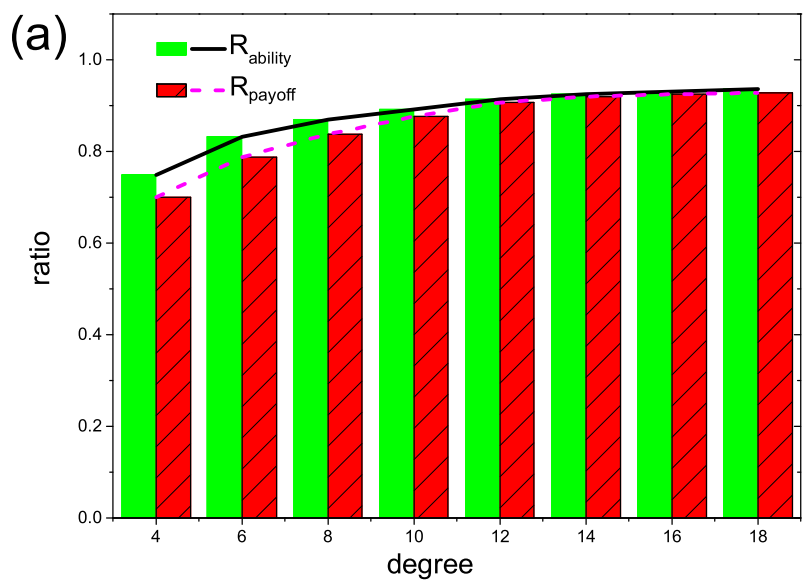
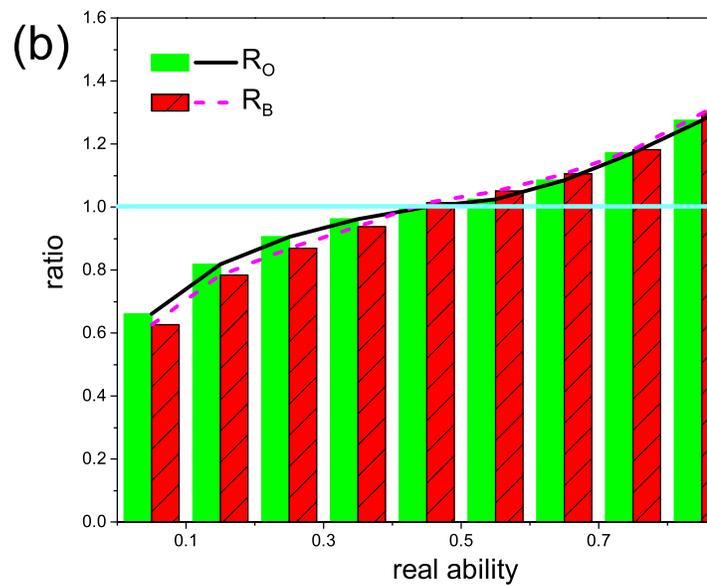
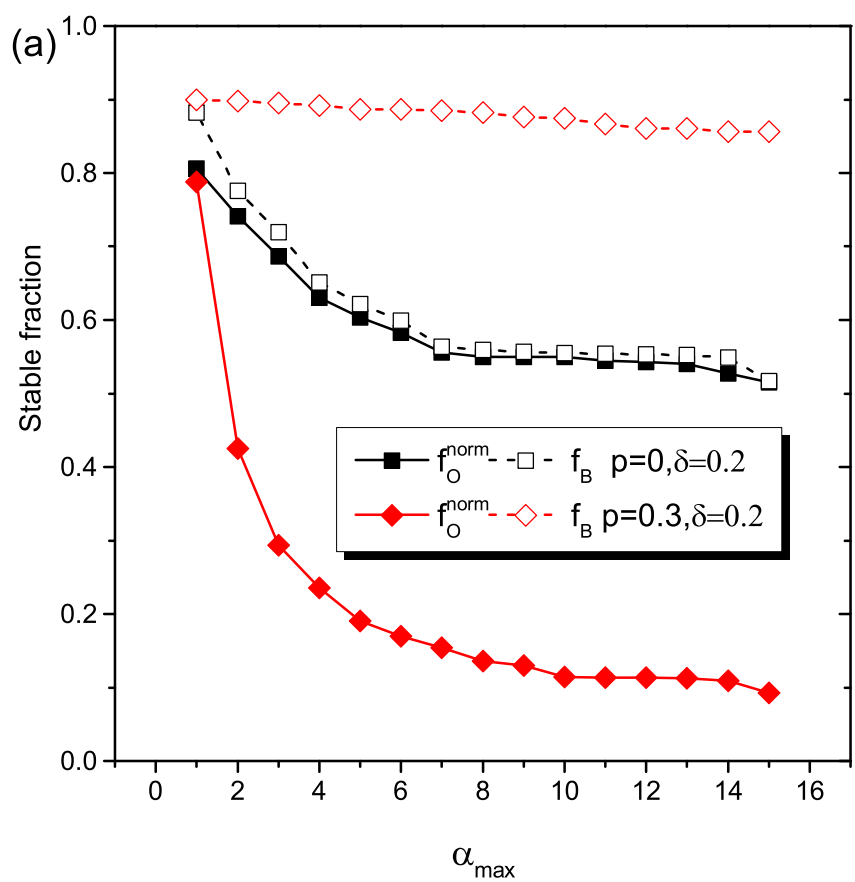
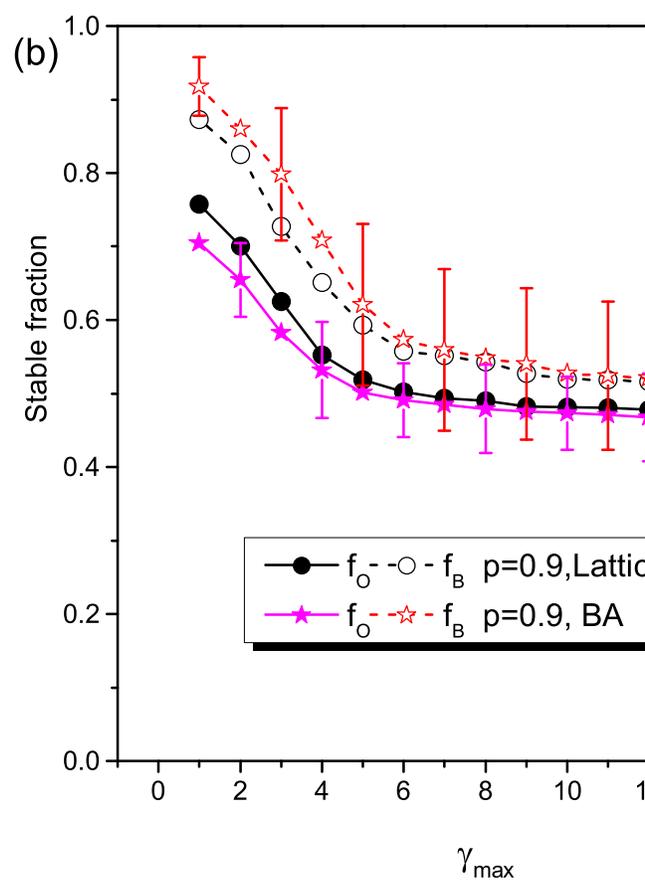
\end{document}